\documentclass[aps,preprint,nofootinbib]{revtex4}
\usepackage{amssymb}
\usepackage{graphicx}
\usepackage{epsfig}

\begin{document}

%=====================================
%       put your definitions here
%=====================================
\def\beq{\begin{eqnarray}}
\def\eeq{\end{eqnarray}}
\def\non{\nonumber}
\def\la{\langle}
\def\ra{\rangle}
\def\Un{{\cal U}}
\def\Mbar{\overline{M^0}}
\def\Bmixing{B^0-\overline{B^0}}
\def\Dmixing{D^0-\overline{D^0}}

\def\pr{{Phys. Rev.}~}
\def\prl{{ Phys. Rev. Lett.}~}
\def\pl{{ Phys. Lett.}~}
\def\npb{{ Nucl. Phys. B}~}
\def\epjc{{ Eur. Phys. J. C}~}

%======================================================================

\title{ Probing unparticle theory via lepton flavor violating process
$J/\psi\to ll'$ at BESIII }

\author{Zheng-Tao Wei}
\author{Ye Xu}
\author{Xue-Qian Li}

\affiliation{ Department of Physics, Nankai University,
  Tianjin 300071, China}

\begin{abstract}

\noindent The lepton flavor violating process $J/\psi\rightarrow
ll'\;\; (l\neq l')$ serves as an ideal place to probe the unparticle
theory. Such process can only occur at loop level in the Standard
model (SM), so that should be very suppressed, by contrast in
unparticle scenario, it happens at tree level and its contribution
may be sizable for practical measurement. Moreover, the BESIII will
offer the largest database on $J/\psi$ which makes more accurate
measurements possible. Furthermore, for such purely leptonic decays
background is relatively low and signal would be cleaner. Our work
carefully investigates the possibility of observing such processes
from both theoretical and experimental aspects.

\end{abstract}

\maketitle

\section{Introduction}

Violation of the lepton flavor symmetry plays an important role for
exploring new physics beyond the standard model (SM). Neutrino
oscillations which are observed in many experiments, indicate that
lepton flavor symmetry is broken and the minimal SM must be extended
to accommodate non-zero neutrino masses. Up to present, the lepton
flavor violation (LFV) has only been observed in the neutrino
sector. This violation is induced by the neutrino masses and
existence of the right-handed neutrinos. By contrast, for the
charged lepton sector, there is no any evidence for such a
violation. Many studies have been devoted to explore the LFV through
$\mu$, $\tau$ and $Z$ decays \cite{Feng}. In Ref. \cite{Zhang}, it
was suggested to look for LFV via charmonium $J/\psi$ decays. In
terms of the database of $5.8\times 10^7$ $J/\psi$ data at BES
processes $J/\psi\to e\mu$ \cite{BES1}, $J/\psi\to \mu\tau,~e\tau$
\cite{BES2} have been explored and upper limits for their branching
fractions were set. New hope is raised that the forthcoming upgraded
BESIII will accumulate up to $10^{10}$ $J/\psi$ per year
\cite{BESIII}, which makes it possible to test LFV with a much
higher precision. It is well known that the SM contribution to such
violation can only occur via loops and are very suppressed, so that
cannot produce observable effects even with so large database of
$J/\psi$. Therefore it is time to investigate possibility of
discovering such violation in terms of the theoretical models beyond
SM.

Recently, Georgi proposed an interesting idea that a scale invariant
stuff may exist in our world. The scale invariant stuff contains no
particle, but the so-called unparticle \cite{Georgi}. To be
consistent with the present experimental observations, in the
effective theory about the unparticle, the coupling of unparticle to
the ordinary Standard Model (SM) matter must be sufficiently weak.
The scale dimension of unparticle is in general fractional rather
than an integral number. The interactions between the unparticle and
the SM particles in low energy effective theory can lead to various
interesting phenomenology. There have been many phenomenological and
theoretical explorations related to unparticles in literature. Since
interaction between unparticle and ordinary SM matter is rather
weak, we can expect that its participation cannot produce observable
effects for the processes (production and decays) which have large
rates, but definitely may play important roles in the rare processes
where the SM contribution are suppressed.

In unparticle physics, the unparticle can couple to different
flavors of the SM leptons and thus induce LFV even at tree level.
There have been several studies of LFV in unparticle physics, for
instance, $\mu\to 3e$ in \cite{ACG}; $\mu\to e+\Un,~3e$ in
\cite{CGM};  $M^0\to ll'$, $e^+e^-\to ll'$ in \cite{LWW}; $\mu\to
e\gamma$, $\mu-e$ conversion in \cite{DY}; muonium and antimuonium
oscillation in \cite{CHLTW}; $r\to ll'$, $l\to l'\gamma\gamma$ in
\cite{IITan}. In this work, we propose to study LFV in unparticle
physics via $J/\psi\to ll'$ decays. There are several advantages of
exploring LFV in $J/\psi\to ll'$ decays, in particular the
$J/\psi\to e\mu$ process. On the experimental side, the final states
contain only leptons which are easy to detect. Compared to
$e^+e^-\to ll'$ processes which occur due to non-resonance
contribution, a large database of $J/\psi$ is expected at BESIII
which will begin running very soon. From the theoretical point of
view, the processes $J/\psi\to ll'$ are free of hadronic
uncertainties after the parameter related to the binding effect of
$J/\psi$ is experimentally determined by measuring $J/\psi\to
e^+e^-(\mu^+\mu^-)$.

Anyhow, an observation of $J/\psi\to ll'\; (l\neq l')$ would be a
clear indication of new physics beyond SM. In our case, $c\bar c$ in
$J/\psi$ annihilate into a virtual unparticle which later converts
into two leptons $ll'$ with different flavors. As we will show that
the theoretical prediction on the branching ratio of $J/\psi\to
e\mu$ can reach about $10^{-8}$ order, which is absolutely possible
to be observed at BESIII. To help or experimental colleagues, on the
basis of BESIII environments, we carry out an investigation on the
possibility of search for $J/\psi\to e\mu$ in terms of a Monta Carlo
simulation.

\section{ $J/\psi\to ll'$ in unparticle physics}

We start with a brief review about the unparticle physics which is
relevant to this study. The scale invariant unparticle fields emerge
below an energy scale $\Lambda_\Un$ which is assumed to be at the
order of TeV. The interactions of the unparticle with the SM
particles are described by a low energy effective theory. The
coupling of unparticle to SM fermions (quarks and leptons) is
generally given by the following effective operators as
 \beq
 \frac{c_{SS}^{f'f}}{\Lambda_\Un^{d_\Un-1}}\bar f'f O_\Un,  \qquad
 \frac{c_{SP}^{f'f}}{\Lambda_\Un^{d_\Un-1}}\bar f'\gamma_5 f O_\Un, \qquad
 \frac{c_{SV}^{f'f}}{\Lambda_\Un^{d_\Un}}\bar f'\gamma_{\mu}f \partial^\mu O_\Un,  \qquad
 \frac{c_{SA}^{f'f}}{\Lambda_\Un^{d_\Un}}\bar f'\gamma_{\mu}\gamma_5f \partial^\mu O_\Un; \non\\
 \frac{c_{VS}^{f'f}}{\Lambda_\Un^{d_\Un}}\bar f'f \partial_\mu O_\Un^\mu,  \qquad
 \frac{c_{VP}^{f'f}}{\Lambda_\Un^{d_\Un}}\bar f'\gamma_5 f \partial_\mu O_\Un^\mu, \qquad
 \frac{c_{VV}^{f'f}}{\Lambda_\Un^{d_\Un-1}}\bar f'\gamma_{\mu}f  O_\Un^\mu,  \qquad
 \frac{c_{VA}^{f'f}}{\Lambda_\Un^{d_\Un-1}}\bar f'\gamma_{\mu}\gamma_5f O_\Un^\mu,
 \eeq
where $O_\Un$ and $O_\Un^{\mu}$ denote the scalar and vector
unparticle fields, respectively. Tensor unparticle does not
contribute in our case. The $c_i^j$ are dimensionless coefficients.
Because $J/\psi$ is a vector, only the vector current $\bar
f'\gamma_{\mu}f$ couples to $J/\psi$. For a scalar unparticle
coupling $\bar f'\gamma_{\mu}f
\partial^\mu O_\Un$, it is proportional to the momentum of the unparticle.
Using the equations of motion, one can immediately show that this
part is proportional to the lepton masses in decays of $J/\psi\to
ll'$ and thus is suppressed. Consequently, we will only consider the
effective interaction
$\frac{c_{VV}^{f'f}}{\Lambda_\Un^{d_\Un-1}}\bar f'\gamma_{\mu}f
O_\Un^\mu$ which dominates in the decays of $J/\psi\to ll'$.

For the vector unparticle field, the propagator is given by
 \beq
 \int d^4 x e^{iP\cdot x}\la 0 |TO^{\mu}_\Un(x)O^{\nu}_\Un(0)|0\ra &=&
  i\frac{A_{d_\Un}}{2{\rm sin}(d_\Un\pi)}\frac{-g^{\mu\nu}+P^{\mu}P^{\nu}/P^2}
  {(P^2+i\epsilon)^{2-d_\Un}}e^{-i(d_\Un-2)\pi}.
 \eeq
where
 \beq
 A_{d_\Un}=\frac{16\pi^{5/2}}{(2\pi)^{2d_\Un}}\frac{\Gamma(d_\Un+1/2)}
  {\Gamma(d_\Un-1)\Gamma(2d_\Un)}.
 \eeq
The function ${\rm sin}(d_\Un\pi)$ at the denominator implies that
the scale dimension $d_\Un$ cannot be integers except $d_\Un=1$. The
peculiar unparticle propagator and the phase factor
$e^{-i(d_\Un-2)\pi}$ which provides a CP conserving phase produces
novel effects in high energy scattering processes \cite{CKY}, CP
violation \cite{CG}, neutral meson mixing \cite{neutral} and
neutrino processes \cite{neutrino}.

\begin{figure}[!htb]
\begin{center}
\begin{tabular}{cc}
\includegraphics[width=5cm]{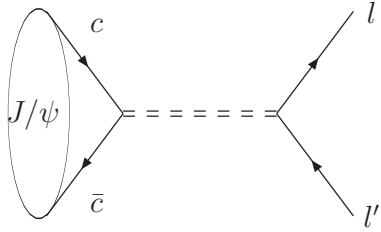}
\end{tabular}
\end{center}
\caption{ The diagram for the decay of $J/\psi\to ll'$. The double
dashed lines represent the unparticle.} \label{fig1}
\end{figure}

The lowest order contribution to decays of $J/\psi\to ll'$ is a
simple tree diagram, which is depicted in Fig. \ref{fig1}. From the
diagram, the exchange of vector unparticle is analogous to a photon
exchange unless photon cannot couple to flavor changing currents. By
using the effective operators and unparticle propagator, it is
straightforward to write out the decay amplitude as
 \beq
 {\cal M}\left(J/\psi(P)\to l(k)l'(k')\right)=\frac{c_{VV}^{cc}c_{VV}^{ll'}}
  {\Lambda_\Un^{2d_\Un-2}}\frac{A_{d_\Un}}{2{\rm sin} d_\Un\pi}\frac{m_{\psi}
  f_{\psi}}{s^{2-d_\Un}}\epsilon^\mu_\psi \bar u(k)\gamma_\mu v(k').
 \eeq
where $s=m_\psi^2$. The decay constant $f_{\psi}$ is derived from
the pure-leptonic process $J/\psi\to e^+e^-(\mu^+\mu^-)$. In
principle, unparticle can also contribute to $J/\psi\to e^+e^-$, but
compared with the QED contribution, its fraction in $BR(J/\psi\to
e^+e^-)\sim 6\%$ and $BR(J/\psi\to e\mu)< 1.1\times 10^{-6}$
\cite{BES1} is completely negligible.

The decay width of $J/\psi\to ll'$ is given by
 \beq
 \Gamma(J/\psi\to l^+l'^-)=\frac{|\vec p_k|}{8\pi
 m_\psi^2}\cdot\frac{1}{3}\sum_{\rm spins}|{\cal M}|^2,
 \eeq
with
 \beq
 \frac{1}{3}\sum_{\rm spins}|{\cal M}|^2=\frac{1}{3}
  \left(\frac{c_{VV}^{cc}c_{VV}^{ll'}}{\Lambda_\Un^{2d_\Un-2}}
  \frac{A_{d_\Un}}{2{\rm sin} d_\Un\pi}\frac{m_{\psi}f_{\psi}}
  {s^{2-d_\Un}}\right)^2 8\Big[ 2(P\cdot k)(P\cdot k')+k\cdot k'
  \Big].
 \eeq

For $J/\psi\to e\mu$, where the final lepton masses are both small
compared to mass of $J/\psi$, we have the simplified form for the
decay ratio as
 \beq \label{eq:jemu}
 \frac{\Gamma(J/\psi\to e^+\mu^-)}{\Gamma(J/\psi\to e^+e^-)}=
  \left|\frac{c_{VV}^{cc}c_{VV}^{e\mu}}{4\pi \alpha e_c}
  \frac{A_{d_\Un}}{2{\rm sin} d_\Un\pi}
  \left(\frac{s}{\Lambda_\Un^2}\right)^{d_\Un-1}\right|^2.
 \eeq
with $e_c=2/3$. Noting that $\Gamma(J/\psi\to
e^-\mu^+)=\Gamma(J/\psi\to e^+\mu^-)$ due to CP symmetry, we have
 \beq
 \Gamma(J/\psi\to e\mu)&\equiv&
 \Gamma(J/\psi\to e^-\mu^+)+\Gamma(J/\psi\to e^+\mu^-) \non\\
 &=&2\Gamma(J/\psi\to e^+\mu^-).
 \eeq

\begin{figure}[!htb]
\begin{center}
\begin{tabular}{cc}
\includegraphics[width=8cm]{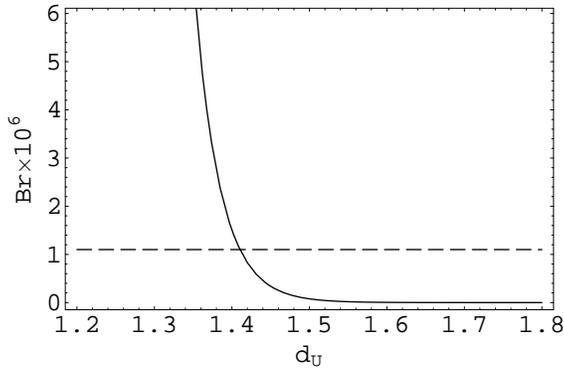}
\end{tabular}
\end{center}
\caption{ The branching ratio for $J/\psi\to e\mu$ with scale
dimension $d_\Un$. The dashed line represents the present
experimental upper bound.} \label{fig2}
\end{figure}

We would like to present several comments as follows: (1) The
relation of Eq. (\ref{eq:jemu}) can be applied to other vector meson
whose mass is larger than that of $J/\psi$, such as $\Upsilon$,
etc.. (2) The decay width of $V\to e\mu$ depends on the invariant
mass squared $s$, and is proportional to
$(s/\Lambda_{\Un}^2)^{d_\Un-1}$. When $d_\Un>1$, the decay width
decreases as a power function. (3) For $1<d_\Un<2$, the decay width
decreases faster as $d_\Un$ increases. The dependence of the
branching ratio of $J/\psi\to e\mu)$ on dimension $d_\Un$ is shown
in Fig. \ref{fig2}.

An important constraint on the unparticle interaction which is
related to $J/\psi\to ll'$ processes comes from the invisible decay
of $J/\psi$. In SM, the invisible final states are neutrinos. The
$J/\psi\to\nu\bar\nu$ decays are highly suppressed and can be
neglected. In unparticle physics, the unparticle $\Un$ cannot be
detected by experimental apparatus and if $J/\psi$ decays into an
unparticle pair, $\Un$'s would correspond to missing energy.
Moreover, the $J/\psi\to\Un$ decay is possible because the mass of
$\Un$ is not fixed, and $\Un$ also escapes from our detector. It
provides a constraint on the the coupling of unparticle to $c\bar c$
quark pair. The rate of the $J/\psi\to\Un$ decay are derived using
the same effective interactions as that in $J/\psi\to ll'$, and it
is straightforwardly to obtain
 \beq
 \frac{\Gamma(J/\psi\to \Un)}{\Gamma(J/\psi\to e^+e^-(\mu^+\mu^-))}=
  \frac{3A_{d_\Un}|c_{VV}^{cc}|^2}{8\pi \alpha^2 e_c^2}
  \left(\frac{s}{\Lambda_\Un^2}\right)^{d_\Un-1}.
 \eeq
Our result is in agreement with \cite{CHT} except a factor of $4$
difference due to different conventions. Another constraint on the
coupling of unparticle to $e\mu$ may come from $\mu\to e+\Un$ and
$\mu\to 3e$. In \cite{CGM}, the scale dimension for vector
unparticle $d_\Un>2$ is used. We will follow \cite{Georgi} and
assume $1<d_\Un<2$. If $d_\Un>2$ as in \cite{CGM}, the branching
ratio of $J/\psi\to e\mu$ will be so small that it cannot be
observed at BESIII only if $d_\Un$ is close to $2$ (other integers
are also possible).

The BESII collaboration recently sets an upper limit for the
invisible decay ratio $\frac{\Gamma(J/\psi\to \Un)}{\Gamma(J/\psi\to
\mu^+\mu^-)}<1.2\times 10^{-2}$ \cite{BESinv}. It determines the
upper bound on the coupling coefficient of unparticle to $c\bar c$
quarks. If choosing $\Lambda_\Un=1$ TeV and $d_\Un=1.35$, then we
obtain $|c^{cc}_{VV}|<0.01$.

The present experimental upper bounds on the branching fractions are
given in \cite{BES1,BES2}: $BR(J/\psi\to e\mu)<1.1\times 10^{-6}$,
$BR(J/\psi\to e\tau)<2.0\times 10^{-6}$, $BR(J/\psi\to
\mu\tau)<8.3\times 10^{-6}$. Among them, the upper bound on
$J/\psi\to e\mu$ is the lowest which may provide the most stringent
constraint on the coupling. If we adopt  the constraints from
invisible decay to fix the unparticle parameters, i.e. $d_\Un=1.35$,
$\Lambda_\Un=1$ TeV and choose $c_{VV}^{cc}c_{VV}^{e\mu}=0.01$, the
branching fraction is predicted to be $BR(J/\psi\to e\mu)=7\times
10^{-8}$. We may further constrain the unparticle parameters by the
upper bound on $BR(J/\psi\to e\mu)$. When $\Lambda_\Un$ and
$c_{VV}^{cc}c_{VV}^{e\mu}$ are fixed, for instance, choosing
$\Lambda_\Un=1$ TeV and $c_{VV}^{cc}c_{VV}^{e\mu}=0.01$, we obtain
$d_\Un>1.26$. This result is consistent with the constraint achieved
from $D^0-\bar D^0$ mixing within $1\sigma$ error tolerance
\cite{CHLTW}: $d_\Un=(0.21+n)\pm 0.07$ where $n$ is an arbitrary
integer (note that uncertainty is still large). We will discuss the
possibility of observing fractions of order $10^{-8}$ at BESIII in
the next section.

In \cite{Zhang}, the authors obtain a bound $BR(J/\psi\to
e\mu)<8.3\times 10^{-6}$ based on a model-independent approach which
introduces an effective four-fermion contact interaction. By
considering the strong bounds from three body LFV processes, such as
$\mu\to 3e$ and using the "unitarity inspired" relation between two
and three-body LFV decays, it is found that the upper bound is
constrained to be very small $BR(J/\psi\to e\mu)<4\times 10^{-13}$
\cite{Nussinov:2000nm}. However, it is noted that such serious bound
can be avoided due to some possible kinematical suppression or
cancelations \cite{Zhang}. Our predicted value lies a range between
\cite{Zhang} and \cite{Nussinov:2000nm}. The precise prediction is
difficult because of the unknown model parameters, especially the
scale dimension \footnote{For scale dimension of the vector
unparticle, it is pointed out that conformal symmetry puts a lower
bound $d\geq 3$ in \cite{Grinstein:2008qk}. If we take this
constraint, the fraction of $BR(J/\psi\to ll')$ will be decreased by
more than two orders and it is impossible to observe them in
experiment except $d_\Un$ is very close to 3. One may avoid this
problem by taking a point of view that scale symmetry is different
from conformal symmetry. The discussion of this controversial topic
is beyond this study.}.

\section{Estimating Possibility of the Determination of $BR(J/\psi \to e\mu)$}

BESIII is a spectrometer which will begin operation in 2008 at the
upgraded Beijing Electron-Positron Collider (BEPCII) whose designed
luminosity is approximately 100 times higher than that of BEPC
\cite{BESIII}. With this high luminosity, the BESIII detector will
be able to collect over 10 billion $J/\psi$ events in one year of
running. The performance of the BESIII  \cite{BESIII} will be much
more effective in comparison with the BESII  \cite{BBB}, so that we
can expect that BESIII would perform a more precise and clear
particle identification (PID)  than BESII. It helps to accurately
analyze rare decay channels, such as the concerned lepton-flavor
violating processes.

The upper bound on $BR(J/\psi \to e\mu)$ was achieved with the the
BESII data \cite{BES1} a long while ago. Before  BESIII begins
running, we would roughly pre-estimate the upper bound with the
BESIII data, if signals are still not observed. For BESIII, ten
billion $J/\psi$ events can be used for our analysis. The main
backgrounds to our signal channel $J/\psi \to e \mu$ are the nearly
back-to-back tracks in $J/\psi\to
e^+e^-,~\mu^+\mu^-,~K^+K^-,~\pi^+\pi^-$ and $e^+e^-\to
e^+e^-(\gamma),~\mu^+\mu^-(\gamma)$. Since the BESIII experiment has
not begun running yet, some detailed information about BESIII
environment is not available, therefore we adopt relevant parameters
for BESII  to estimate $BR(J/\psi \to e\mu)$ and the corresponding
backgrounds, except the number of available $J/\psi$ events.

The method of determining the selection efficiency for $J/\psi \to e
\mu$ and the rate of misidentifying the background  as a signal is
the same as the one used in \cite{BES1}. The overall selection
efficiency for the signal channel (or rate of misidentifying the
background channel as the signal channel) includes the $\mu$ and $e$
particle identification (PID) efficiency and the geometric
efficiency. Indeed, the geometric efficiency of BESIII is only
slightly different from BESII, thus in this work we set the same
geometric efficiency as that for the previous BESII experiments. The
PID efficiency and contamination rate which refer to identifying
other particles as electrons or muons in BESIII are taken from
\cite{BES1}. In order to discriminate muons to reject other
particles as much as possible, a very rigorous criterion of the
$\mu$ identification has to be performed while selecting signal
event $J/\psi \to e \mu$. The geometric and the PID efficiency are
listed in Tables \ref{tab1} and \ref{tab2} respectively.

\begin{table}
\begin{center}
 \caption{Monte Carlo geometric efficiency. \label{tab1} }
\begin{tabular}{lcc}\hline\hline
~~~~channel   &    & ~~~MC Efficiency~~~
\\\hline
$J/\psi \to e\mu$   & $\epsilon_{e\mu-MC}$  & 53.7$\%$      \\
$J/\psi \to ee$     & $\epsilon_{ee-MC}$    & 61.47$\%$     \\
$J/\psi \to \mu\mu$ & $\epsilon_{\mu\mu-MC}$& 58.32$\%$     \\
$J/\psi \to \pi\pi$ &$\epsilon_{\pi\pi-MC}$ & 52.74$\%$     \\
$J/\psi \to KK$     & $\epsilon_{KK-MC}$    & 24.38$\%$     \\
$e^+e^- \to e^+e^-(\gamma)$~~~~~~ & $\epsilon_{ee(\gamma)-MC}$
                                            & 32.51$\%$     \\
$e^+e^-\to \mu^+\mu^-(\gamma)$~~~~~~  & $\epsilon_{\mu
\mu(\gamma)-MC}$ &42.96$\%$      \\\hline\hline
\end{tabular}
\end{center}
\end{table}

\begin{table}[htbp]
\begin{center}
\renewcommand{\tabcolsep}{0.3pc}
 \caption{The particle identification/misidentification efficiency. \label{tab2} }
\begin{tabular}{|c|c|c|}\hline\hline
     & ~~identified as $e$~~   & ~~identified as $\mu$~~ \\\hline
    ~~~$e$~~~ & 98.0$\%$          & \hspace{0.5cm}|   \\
    $\mu$     & \hspace{0.5cm}|   & 19.0$\%$          \\
    $\pi$     & 0.1$\%$           & 0.46$\%$          \\
    $K$       & 0.1$\%$           & 0.38$\%$          \\\hline\hline
\end{tabular}
\end{center}
\end{table}
The overall selection efficiency for the signal $J/\psi \to e \mu$
(or rate of misidentifying the background as the signal $J/\psi \to
e \mu$) can be calculated with the PID (or misidentification) and
geometric efficiency given in the tables. For the signal channel
$J/\psi \to e \mu$, the total selection efficiency is then
 \beq
 \epsilon_{J/\psi \to e \mu} = \epsilon_{e \mu-MC} \times \epsilon_{e
  \to e} \times \epsilon_{\mu \to \mu},
 \eeq
where $\epsilon_{e \mu-MC}$ is the geometric efficiency,
$\epsilon_{e \to e}$ is the electron PID efficiency, and
$\epsilon_{\mu \to \mu}$ is the muon PID efficiency. With the number
given in Tables \ref{tab1} and \ref{tab2}, we obtain the selection
efficiency for $J/\psi \to e\mu$ as about 10\%. One can evaluate the
misidentification rate of the background channels $J/\psi \to XX$ or
$e^+e^- \to XX$, where $X=e$, $\mu$, $\pi$, $K$, (i.e.  the
background channels are misidentified as the signal channel) as
 \beq
 \epsilon_{XX} = \epsilon_{XX-MC} \times \epsilon_{X \to e} \times
 \epsilon_{X \to \mu}\times 2.
 \eeq
where $\epsilon_{XX-MC}$ is the Monte-Carlo geometric efficiency,
shown in Table~\ref{tab1}, and $\epsilon_{X \to e}$ and $\epsilon_{X
\to \mu}$ are the contamination rate for X being identified as an
electron or a muon.

Because electrons and muons cannot be misidentified with each other,
the only background which should be taken into account in the
estimation is the hadronic channels. The misidentification rate and
the number of background from hadronic channels are estimated in
terms of the methods given in \cite{BES1}, and the results are
listed in Table \ref{tab3}.
\begin{table}[htbp]
\begin{center}
\caption{The misidentification rate and the number of background
from hadronic channel. \label{tab3} }
\begin{tabular}{|c|c|c|}\hline\hline
 ~~~~~decays~~~~~  &~~misidentification~~ &~~~~number of~~~~\\
 &rate &  background\\\hline
 J/$\psi \to \pi\pi$ &$4.85 \times 10^{-6}$ &7.13\\
 J/$\psi \to KK$ &$1.85 \times 10^{-6}$ &4.38 \\ \hline
  total && 11.5 \\\hline\hline
\end{tabular}
\end{center}
\end{table}

Now, we can set an upper limit for $BR(J/\psi \to e\mu)$ if there
only few events are observed in BESIII and they are consistent with
the background estimation. For an illustration, we discuss a case
similar to that we did for BESII where there were four $J/\psi \to
e\mu$ candidates observed by the detector \cite{BES1}. Although our
real case may be different, the analysis should provide a useful
reference. With the four observed events which are supposed to be
the signal and 11.5 background events from hadronic channels in
BESIII, we gave an upper limit for $BR(J/\psi \to e\mu)$ as
 \beq
  BR<\lambda(N_{OB},N_{BG})/[N_T \times \epsilon_{J/\Psi \to e\mu}].
 \eeq
where $\lambda$ is calculated using the method of \cite{JM} to be
$\lambda$=2.31 is obtained at  90$\%$ C.L.. Therefore, we obtain
$BR(J/\Psi \to e\mu)<2.3 \times 10^{-9}$ in the BESIII experiment,
which is more precise than the BESII by two orders.

Now let us discuss the case with branching ratio being about
$BR(J/\psi\to e\mu)=7\times 10^{-8}$ as the unparticle physics
predicts. If the total selection efficiency for $J/\psi \to e\mu$ is
$10\%$,  then, about 70 events are passed through the selection
criterion of $J/\psi\to e\mu$ in 10 billion $J/\psi$ events
collected by BESIII during one year of running. From Table 3, there
are 11.5 background events as estimated. Then, the signal to noise
ratio reaches about 7:1. It is possible to determinate the branching
ratio of $J/\psi\to e\mu$ since the signal significance is enough.

In fact, the electron and muon identification is very important in
the analysis of the concerned  channels. The muon identification in
BESIII is better than the one in BESII. Certainly, the contamination
rate for kaons being misidentified as muons should be lower than the
one for pions \cite{BBB}. In the work, however, the contamination
rate for kaons being misidentified as muons has to be set to the
same as the contamination rate for pions being misidentified as
muons, because of lack of precise information about kaons. So the
background for $J/\psi\to e\mu$ is overestimated in this work. In
fact, the background in BESIII should be less than that estimated in
the work. Therefore, the real signal to noise ratio for $J/\psi\to
e\mu$ is higher in real BESIII experiments.

\section{Conclusions}

The lepton flavor violating decays of $J/\psi\to ll'$ provide an
ideal probe to explore new physics beyond the minimal standard
model. In this study, we have presented a detailed analysis on the
contribution of unparticle physics to $J/\psi\to ll'$ decays. The
coupling of unparticle to different flavors of leptons cause the LFV
processes at tree level.

Even though the scaling factor
$(m_{\psi}^2/\Lambda_\Un^2)^{d_\Un-1}$ which is sensitive to the
scale dimension of the unparticle field suppresses LFV, one still
hopes to observe such rare processes at experiments with very high
luminosity and precision. Within a reasonably chosen parameter
space, the branching fraction of $J/\psi\to e\mu$ can reach about an
order of $10^{-8}$ which is possible to be observed at the
forthcoming BESIII.

Indeed, the BESIII is planning to produce $10^{10}$ $J/\psi$ per
year. These huge data samples guarantee an exploration of LFV with a
high precision. As a moderate estimation, the upper limit of
measuring $BR(J/\psi\to e\mu)$ is obtained as $2\times 10^{-9}$ at
BESIII. If so, the unparticle theory indeed predicts a production
rate within this range and LFV should be observed. As discussed
above, the background for such lepton-flavor violating processes is
relatively clean and the signal-to-noise is high, so that
observation would be more optimistic.

As we know, to judge validity of a new physics model, usually one
cannot decide its existence only by a unique experiment, but needs a
combination of data from various measurements. Such a synthesis
analysis may help to confirm a new model. Phenomenological analysis
on effects induced by unparticle physics has been carried out for
various production and decay processes and even the cosmology. In
general, the concerned parameters, such as its coupling to SM
matter, scale dimension $d_\Un$ and other characteristics are
constrained by earlier analysis, therefore it seems to be the time
to let experiment tell us the validity of the whole theory. Thus our
conclusion is that observation of the lepton-flavor violating decays
may provide an important test of the unparticle scenario. If the
observation confirms the predicted value, the unparticle theory
would be greatly supported (even though relatively large uncertainty
still remains), by contrary, if the reaction is not seen at BESIII,
one may set a more stringent constraint on the parameters which
would be applied to other experiments for further tests.

\section*{Acknowledgments}

This work was supported in part by National Natural Science
Foundation of China (NSFC) under contract Nos. 10475042, 10745002,
10705015, 10605014 and the special foundation of the Education
Ministry of China.

\end{document}